\documentclass[aip,reprint,graphicx]{revtex4-1}

\usepackage[english]{babel}
\usepackage[a4paper, top=2cm, bottom=2cm, left=2cm, right=2cm]{geometry}
\usepackage[dvipsnames]{xcolor}
\usepackage{soul}

\usepackage{color}
\usepackage{dcolumn}
\usepackage{amsmath}
\usepackage{dsfont}
\usepackage{epsfig}

\graphicspath{{./}{./figures/}}

\begin{document}

\title{Nonlinear waves at the free surface of flexible mechanical metamaterials} 

\author{Bolei DENG}
\affiliation{Harvard John A. Paulson School of Engineering and Applied Sciences, Harvard University, Cambridge, MA 02138}
\affiliation{Computer Science and Artificial Intelligence Laboratory, Massachusetts Institute of Technology, Cambridge, MA 02139}
\affiliation{Department of Mechanical Engineering, Massachusetts Institute of Technology, Cambridge, MA 02139}

\author{Hang SHU}
\affiliation{Department of Mechanical Engineering and Applied Mechanics, University of Pennsylvania, Philadelphia, PA 19104}

\author{Jian LI}
\affiliation{Harvard John A. Paulson School of Engineering and Applied Sciences, Harvard University, Cambridge, MA 02138}
\affiliation{Department of Engineering Mechanics, Zhejiang University, Hangzhou, 310027, China}

\author{Chengyang MO}
\affiliation{Department of Mechanical Engineering and Applied Mechanics, University of Pennsylvania, Philadelphia, PA 19104}

\author{Jordan R. RANEY}
\affiliation{Department of Mechanical Engineering and Applied Mechanics, University of Pennsylvania, Philadelphia, PA 19104}

\author{Vincent TOURNAT}
\affiliation{Harvard John A. Paulson School of Engineering and Applied Sciences, Harvard University, Cambridge, MA 02138}
\affiliation{Laboratoire d'Acoustique de l'Universit\'e du Mans (LAUM), UMR 6613, Institut d'Acoustique - Graduate School (IA-GS), CNRS, Le Mans Universit\'e, France}

\author{Katia BERTOLDI}
\affiliation{Harvard John A. Paulson School of Engineering and Applied Sciences, Harvard University, Cambridge, MA 02138}


\date{\today}

\begin{abstract}
In this letter we investigate the propagation of nonlinear pulses along 
the free surface of flexible metamaterials based on the rotating squares mechanism. While these metamaterials have previously been shown to support the propagation of elastic vector solitons  through their bulk, here we demonstrate  that they can also support   the stable propagation of  nonlinear pulses along their free surface. Further, we show that  the stability of these surface pulses is higher when they minimally interact with the linear dispersive surface modes. Finally, we provide guidelines to select geometries that  minimize such interactions.
\end{abstract}


\maketitle 


Surface waves that propagate along the boundary of a medium play a key role in a variety of natural and man-made systems. Seismic surface waves cause the ground to shake \cite{ben2012seismic} and surface gravity waves can be observed on rivers, lakes and oceans \cite{mehaute1976}. Further, surface ultrasonic waves are harnessed in non-destructive testing to detect cracks or corrosion \cite{dhital2012fully,dwivedi2018advances} and surface acoustic waves are commonly used to realize electronic systems \cite{morgan2010surface,lange2008surface}. It is therefore important to investigate the physics  of surface waves  in order to better control natural events and  to advance technology. 

Ongoing advances in fabrication are enabling the realization of mechanical metamaterials capable of manipulating  elastic waves in unprecedented ways. These have been used to enable the design of waveguides and filters~\cite{Deymier2013}, energy absorbers~\cite{Frenzel_2016}, energy harvesters~\cite{CarraraEtAl2013} and  vibration isolators~\cite{MeiEtAl2012}. They have also provided a powerful platform to investigate and observe surface waves \cite{brule2014experiments,mu2020review} and topologically protected edge modes  \cite{miniaci2018experimental,ni2015topologically,huber2016topological,serra2018observation,fan2019elastic}. While most mechanical metamaterials  operate in the linear regime, it has been recently shown that large deformations and instabilities  can be exploited to manipulate the propagation of finite amplitude elastic waves~\cite{Fraternali_2014,Raney2016,Hwang_2018, yasuda2019origami,Deng_2021,Deng_2017,Deng_2018NM,deng2019focusing,deng2018effect,Yasuda_2020,li2021propagation,Pajunen_2019}. However, to date most studies have focused on nonlinear pulses propagating in the bulk of these flexible metamaterials. The propagation of large amplitude pulses on their free surfaces has received little attention.

\begin{figure}
 \includegraphics[width=8cm]{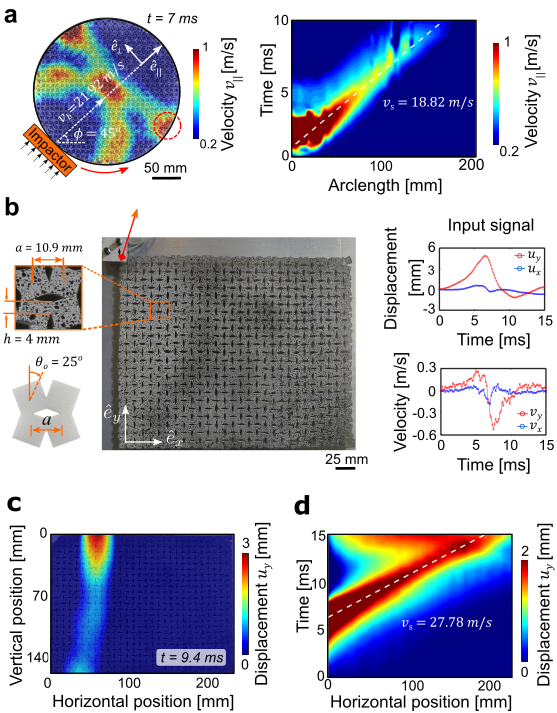}%
\caption{\label{fig1} (a) Experimental snapshot at $t = 7$~ms of a 2D circular sample  with 30 squares along its diameter when
excited with an impactor at $45^\circ$ angle (taken from \citet{deng2019focusing}).  The color represents the  velocity along the direction of the impact ($v_{||}$). The impact also excites a large amplitude pulse  that propagates along the free surface with nearly constant velocity (see red dashed circle in the snapshot and spatio-temporal plot of $v_{||}$ along the surface).  (b) Snapshot of the system tested for this study (left).
Excitation profile of the impacted unit cell (right). (c) Contour plot of the vertical displacement ($u_y$) at time $t = 9.4$~ms after impact.  (d) Spatio-temporal map of   $u_y$ along the top surface. }
\end{figure}

In this letter, we combine experiments and simulations to investigate the propagation of nonlinear waves on the free surface of a flexible metamaterial comprising a network of squares connected by thin and highly deformable ligaments.  Recent studies focused on the propagation of vector solitons through the bulk of such metamaterials have hinted at the existence of large amplitude pulses with stable shape localized at their free surfaces~\cite{deng2019focusing} (Fig.~\ref{fig1}a). Motivated by these observations, we systematically investigate the propagation  of large amplitude waves on the surface of a rectangular sample. We find that the system supports  surface pulses with coupled displacements and rotations that retain their shape during propagation. Further, we  numerically investigate the stability of these surface pulses and find that the less they  interact with the excited linear surface dispersive modes, the more stable they are. 

We consider a $32 \times 24$ array of squares fabricated out of polydimethylsiloxane (PDMS) using direct ink writing~\cite{lewis2006,Deng_2017,deng2019focusing}.  
The squares are rotated by offset angles of $\theta_0=25^\circ$ with a center-to-center-distance of $a = 10.89$~mm, and are connected to one another by flexible ligaments of approximately $4$~mm in width (Fig.~\ref{fig1}b). In our experiments,  we use a customized polylactide (PLA) impactor to apply an impulse to the top left corner of the sample (Fig.~\ref{fig1}b). 
To characterize the propagation of the excited pulses, we record the experiments with a high-speed camera (Photron FASTCAM Mini AX) and 
extract the displacement and velocity  of each square unit. 

Fig.~\ref{fig1}c shows the contour plot of the vertical displacement ($u_y$) at $t=9.4$~ms after impact. The impact excites a pulse with the energy mostly localized close to the top surface. To further characterize the propagation of this pulse, we generate the spatio-temporal map of $u_y$ along the top row of the sample (Fig.~\ref{fig1}d). This indicates that a single pulse is formed and propagates at a speed of $c\simeq 28 $ m/s  until it reaches the end of the specimen. 

\begin{figure*}[htbp]
\includegraphics[width=2\columnwidth]{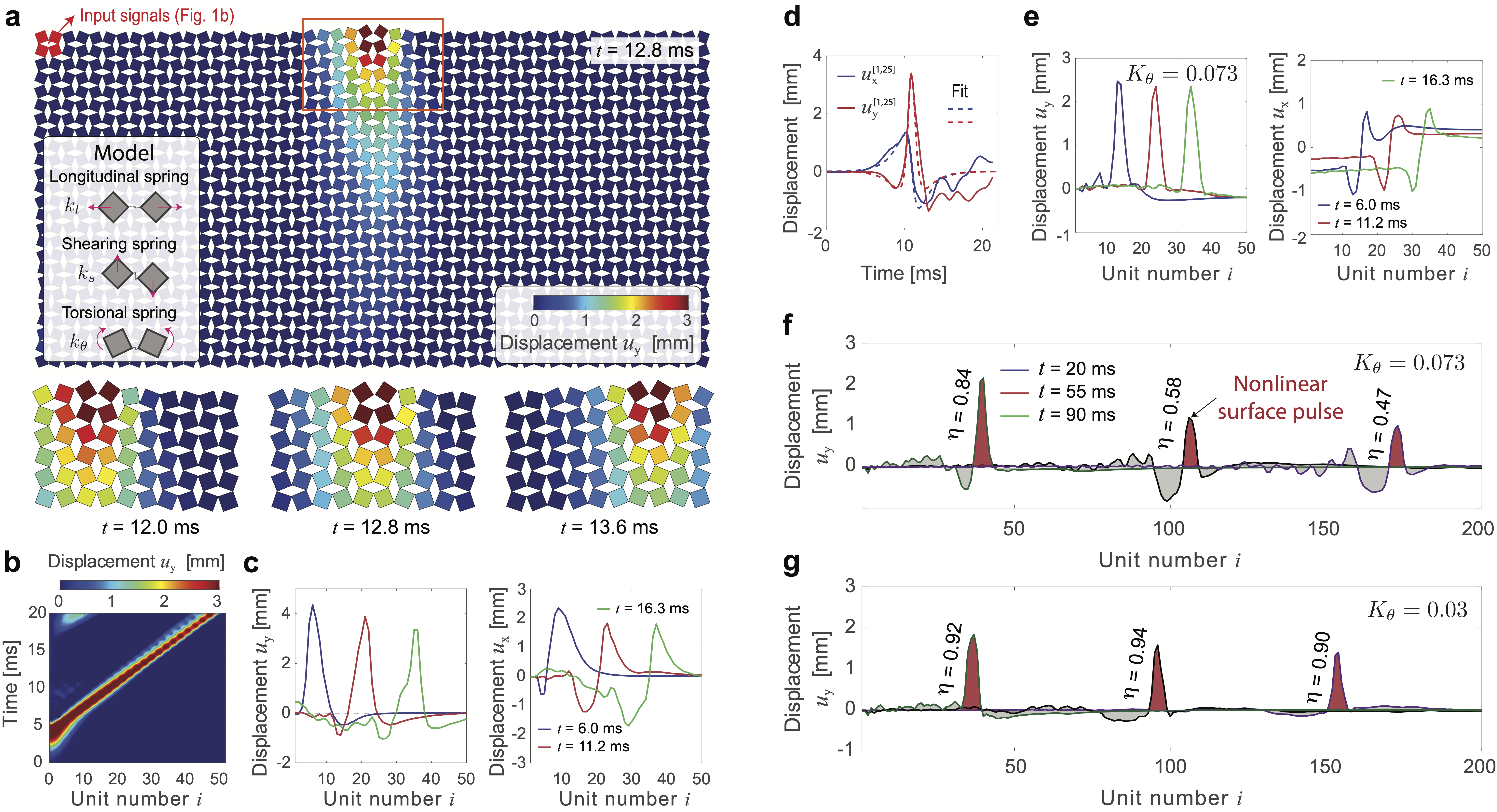}%
\caption{\label{fig2} (a)-(e) Numerical results for a model comprising 50$\times$25 squares.  (a) Contour plots of $u_y$  over the entire model at $t = 12.8$ ms after impact and in a region close to the top surface at $t = 12.0$ ms, 12.8 ms, and 13.6 ms. The pulse is excited by applying the experimentally extracted displacement signal shown in Fig.~\ref{fig1}b  to  the four squares highlighted in red.  (b) Spatio-temporal map of   $u_y$ along the top surface. (c) Spatial displacement profiles along the top surface at $t=$ 6.0, 11.2 and 16.3 ms. (d) The signal collected at the 25$^{th}$ unit (continuous lines) is fitted with derivatives of Gaussian functions (dashed lines). (e) Spatial displacement profiles along the top surface at $t=$ 6.0, 11.2 and 16.3 ms when the model is excited by applying  the
 signal shown in (b). (f)-(g) Numerical results for a model comprising 200$\times$50 squares. Spatial displacement profiles along the top surface at $t=$ 20.0, 55.0 and 90.0 ms  with (f) $K_\theta=0.073$ and (g) $K_\theta=0.03$ when  excited by applying  the
 signal shown in (d). }
\end{figure*}

Next, we make use of numerical simulations to systematically explore the characteristics of the nonlinear pulses that propagate along the surface of the metamaterial. We model the system as an array of rigid squares with  mass $m = 0.4$~g and moment of inertia $J = 4.8$~g~mm$^2$. Each square has  three degrees of freedom (displacements $u_x$ and $u_y$ and rotation $\theta$) and is connected to the neighbors via a combination of linear longitudinal (with experimentally measured stiffness $k_l = 19237 $ N/m), shear ($k_s = 9618 $ N/m), and rotational springs ($k_\theta = 0.0507 $ N.m/rad)\cite{deng2019focusing}. 
By imposing force equilibrium at each unit, we derive a system of coupled nonlinear  ordinary differential equations that we numerically integrate to obtain the response of the structure \cite{deng2018effect}. In our simulations we consider a larger system  comprising  50$\times$25 squares to  minimize  boundary effects, apply the experimentally extracted displacement signal shown in Fig.~\ref{fig1}b  to  four squares on the top left corner (highlighted in red in Fig.~\ref{fig2}a), and implement free-boundary conditions everywhere else. Finally, to prevent reflections from the bottom surface, we add progressively increasing damping to the bottom 10 rows of the model.

In Fig.~\ref{fig2}a we report the numerically predicted contour plot of $u_y$ at time $t=12.8$ ms after the impact. We find that the applied input excites  a pulse with a width of $\simeq$5 squares that remains mostly localized on the top surface. This suggests that dispersion should occur, since the wavelength of the wave is comparable to the structure's spatial period. In order to analyze the stability of such large amplitude pulses during propagation, in Fig.~\ref{fig2}b we report the spatial-temporal map of $u_y$ along the top row of the sample. Further, in Fig.~\ref{fig2}c we show the evolution of the vertical ($u_y$) and  horizontal ($u_x$) displacement components  of the pulse as a function of space along the top surface  at $t=6, 11.2$ and $16.3$ ms. While the former indicates that the pulse travels along the surface with a relatively constant velocity and width, the latter shows that its amplitude and shape vary during propagation. Since such variation could be due to an applied impact that results in a displacement signal far from that of a potentially supported solitary wave, we then use the numerical signal collected at the 25$^{th}$ unit (which we fit with derivatives of Gaussian functions - Fig.~\ref{fig2}d) as new impact signals for both the $u_x$ and $u_y$ components. As shown in Fig.~\ref{fig2}e,  this input  initially results in a more stable propagation along the surface, closer to what one would expect from a solitary wave.  However,  when simulating a longer sample comprising 200$\times$25 units we find that the pulse gets largely distorted after a propagation distance of $\approx$100 units (Fig.~\ref{fig2}f) - likely because of interactions with the linear surface waves. To better quantify this distortion, we introduce the ratio 
\begin{equation}\label{distort}
    \eta(t) =  \frac{\sum_{i\in Set_{p}}\,[u_y^{[1,i]}(t)]^2}{\sum_{i=1}^{200}\,[u_y^{[1,i]}(t)]^2} ,
\end{equation}
where $u_y^{[1,i]}(t)$ is the displacement of the $i$-th square on the top surface along the $y$-direction at time $t$ and $Set_p$ denotes the set of squares on the top surface that are in the nonlinear pulse. This set comprises the squares for which $x^{[1,i]}\in[x_0-3W,x_0+3W]$, where $x_0$ and $W$ denote the position and width of the nonlinear pulse, which are identified by fitting $u_y^{[1,i]}$ with a Gaussian curve ($A\,\mbox{sech}((x-x_0)/W)$).  As shown in Fig.~\ref{fig2}f, we find that $\eta$ is close to 1 at $t$=20 ms, confirming that  the energy is initially  concentrated in the nonlinear pulse. However, during propagation  $\eta$ monotonically decreases ($\eta=0.58$ and 0.47 at $t$=55 and 90 ms, respectively), indicating that the energy progressively leaks out of the  nonlinear pulse. 

Interestingly, our simulations also indicate that the distortion of the pulse  is largely affected by the mechanical properties of the hinges. By changing the torsional stiffness from $K_\theta = 4k_\theta \cos^2\theta_0/(k_l a^2)=0.073$ (Fig.~\ref{fig2}f) to $K_{\theta}=0.03$ (Fig.~\ref{fig2}g), we obtain a surface pulse that seems to be  able to propagate stably with nearly constant shape, amplitude and speed over 200 units. In this case we find that $\eta\approx0.9$ during the entire propagation.

\begin{figure}[htbp]
\includegraphics[width=1\columnwidth]{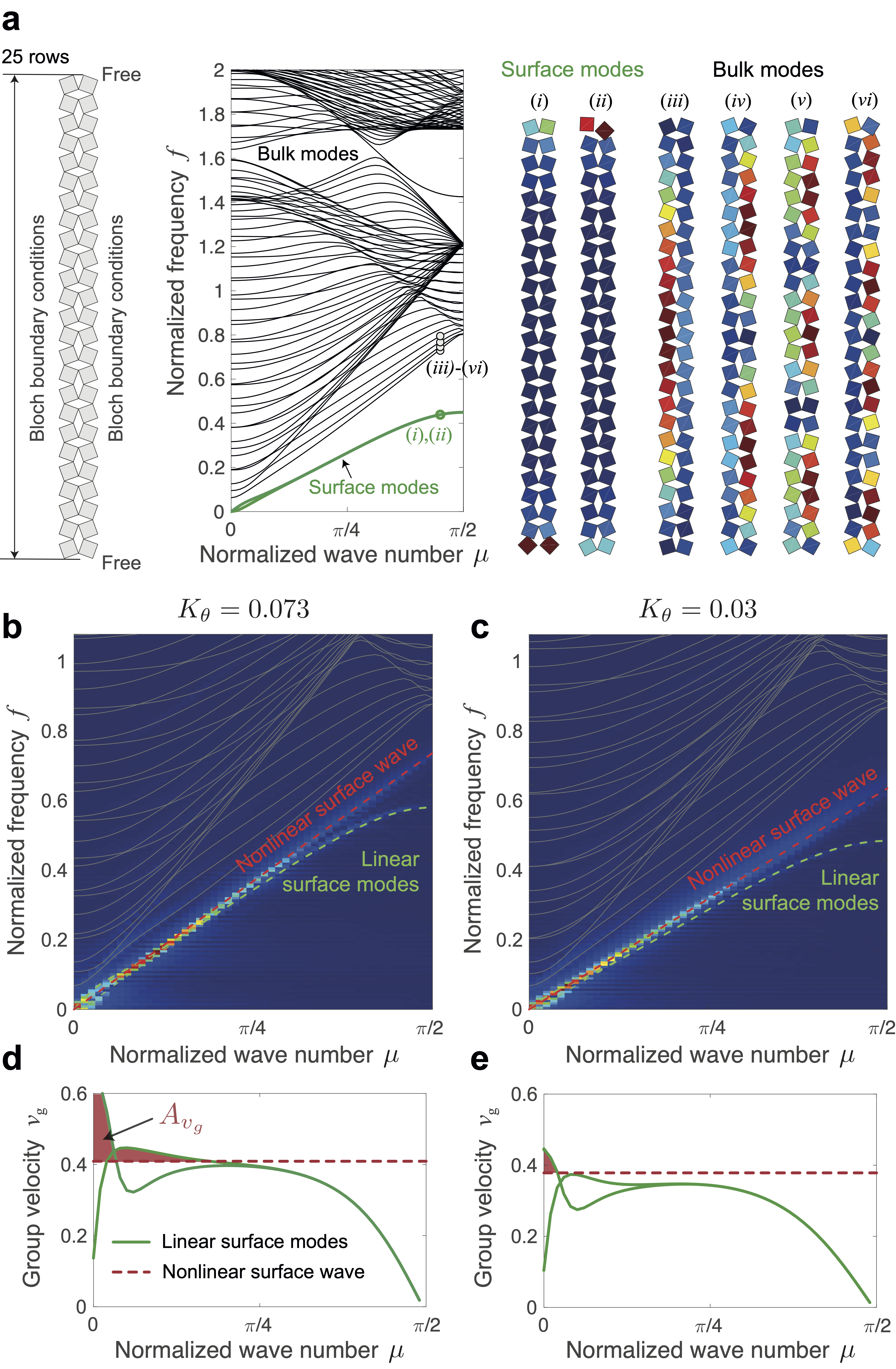}%
\caption{\label{fig3} (a) 1D band structure showing bulk bands (black lines)
and edge bands (green lines). Modal
deformation fields of surfaces ($i$-$ii$) and bulk ($iii$-$vi$) modes. The colors indicate the rotation of each unit (with dark blue corresponding to no rotation and dark red  to maximum rotation). (b)-(c) Comparison between the linear modes and the dispersion curves extracted from
the nonlinear pulses reported in Figs.~\ref{fig2}f and \ref{fig2}g via a double Fourier transform (from space-time to wave number-frequency)  for (b) $K_\theta = 0.073$ and (c) $K_\theta = 0.03$. (d)-(e) Group velocity of the nonlinear (red lines) and linear (green lines) surface waves  for (d) $K_\theta = 0.073$ and (e) $K_\theta = 0.03$.  }
\end{figure}

To verify our hypothesis that the distortion of the nonlinear pulses is caused by interactions with linear surface modes, we calculate the band structure of the system. To this end,  we perform one-dimensional Bloch wave analysis on a supercell comprising 25 $\times$ 2 square units, assuming free boundary conditions for the top and bottom edges. As reported in Fig~\ref{fig3}c for a structure characterized by $K_\theta=0.03$, the band structure shows both bulk modes with motion distributed over the entire supercell (see modes $iii$-$vi$) and surface  modes localized at  the free boundary  (see modes $i$-$ii$). Such surface modes occur at lower frequencies than the bulk modes for comparable wavelengths, and are dispersive. Next, we compare these linear modes to the   dispersion curves extracted from  the nonlinear pulses reported in Figs.~\ref{fig2}f and ~\ref{fig2}g via a double Fourier transform (from space-time to wave number-frequency). We find that  the dispersion curve of the linear and nonlinear surface waves are very close to each other for the metamaterials with  both $K_\theta =0.073$ (Fig.~\ref{fig3}b)  and $K_\theta =0.03$ (Fig.~\ref{fig3}c). However, the nonlinear surface pulses are characterized by a non-dispersive  propagation (i.e. they are a straight line) unlike the dispersion predicted for linear surface modes. This indicates that for the nonlinear pulses the linear dispersion is compensated by nonlinear distortion effects, a clear feature of solitary waves \cite{dauxois2006}.   


In order to quantify the proximity between the linear and nonlinear surface modes, in Figs.~\ref{fig3}d  and ~\ref{fig3}e we report the group velocities   as a function of the wavenumber of the linear (green lines) and nonlinear (red lines) pulses for  $K_{\theta}=0.073$ and 0.03. As expected, we find that the group velocity is constant for the nonlinear modes (i.e. $v_g/v_0 \approx 0.41$ for $K_\theta = 0.073$ and $v_g/v_0 \approx 0.38$ for $K_\theta = 0.03$), whereas  it varies as a function of the wave number for the linear one.  Further, it appears that for the metamaterial with $K_{\theta}=0.03$ the group velocity of the nonlinear pulse is, for most wavenumbers, larger than the one of linear modes (Fig.~\ref{fig3}d), ensuring separation and weak interactions between them. 
By contrast, for the structure with $K_{\theta}=0.073$  the group velocity of the linear waves is larger than that of the nonlinear pulse over a wider range of wavenumbers (see area highlighted in red in Fig.~\ref{fig3}e). It follows that in this case the linear waves propagate faster than the nonlinear pulse for a wide range of wavenumbers and this promotes  interactions between them that ultimately lead to distortion of  the nonlinear pulse during propagation. 
To quantify such interactions, we calculate the area $A_{v_g}$ of the region below $v_g$ of the linear surface modes, but above that of the nonlinear pulse (see regions highlighted in red in Figs.~\ref{fig3}d and \ref{fig3}e). For the two structures with  $K_\theta = 0.073$ and $K_\theta = 0.03$ we find that $A_{v_g}=0.15$ and 0.009, respectively.

\begin{figure*}[tbh!]
\includegraphics[width=2\columnwidth]{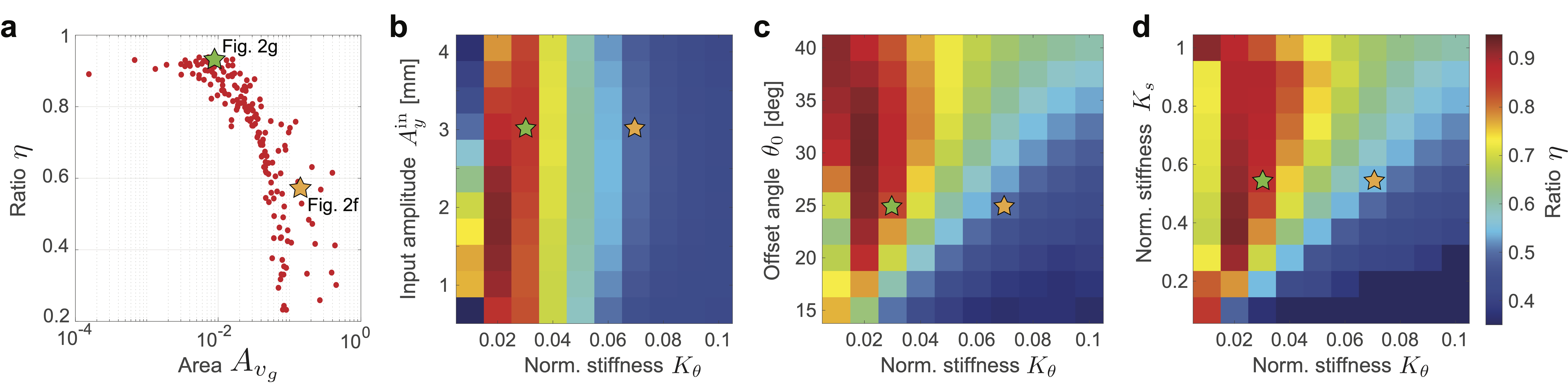}%
\caption{\label{fig4} (a) Relation between   $\eta$ and   $A_{v_g}$ for the 330 simulated metamaterials.   (b)-(d) Evolution of $\eta$ as a function of $K_\theta$, $K_s$, input amplitude $A_\text{in}$ and offet angle $\theta_0$. The green and yellow stars correspond to  the two structures considered in Figs.~\ref{fig2}f and \ref{fig2}g, respectively. }
\end{figure*}


Finally, to confirm the connection between the distortion of the nonlinear pulses and proximity between the linear and nonlinear surface modes, we simulate 330 systems characterized by $K_\theta \in[0.01,0.1]$, $K_s=k_s/k_l$ $\in[0.1,1]$, $\theta_0\in[15^\circ,40^\circ]$ and input amplitude $A\in[1 \text{mm},4 \text{mm}]$. From each simulation, we extract the mean value of $\eta$ (defined in Eq.~(\ref{distort}) and averaged over 10 values calculated at 10 times between 55 ms and 90 ms), as well $A_{v_g}$.  As shown in Fig.~\ref{fig4}a, we find that the smaller $A_{v_g}$ (i.e., the more separation there is between the group velocities of the linear and nonlinear surface pulses), the higher is $\eta$ (i.e., the more energy is  concentrated in the nonlinear pulses). This observation clearly confirms that, for a given metamaterial design, nonlinear pulses are more stable  when they weakly interact  with the linear modes - a condition that is achieved when the group velocity of the nonlinear waves is greater than that of linear modes over a wide frequency range. Further, our results indicate that the ratio $\eta$ strongly depends on the geometric parameters of the  metamaterial. Stable propagation (i.e. $\eta\rightarrow 1$) is found for $K_{\theta} \simeq 0.02$ (Figs.~\ref{fig4}b-d), input amplitude $A^{in} \simeq 2$ mm (Fig.~\ref{fig4}b), offset angle $\theta_0 \simeq 30^o$  (Fig.~\ref{fig4}c),  and dimensionless shear stiffness $K_s=k_s/k_l\simeq 0.5$ (Fig.~\ref{fig4}d). As such, these results provide guidelines to identify flexible metamaterials based on the rotating squares mechanism that can support stable propagation of large amplitude pulses on their free surfaces.

To summarize, we have demonstrated that flexible metamaterials based on the rotating squares mechanism can support the propagation of solitary-like nonlinear wave pulses along their free surfaces. Our results indicate that stable propagation of nonlinear pulses along the surface is achieved when  the large amplitude waves minimally interact with the  linear surface dispersive modes. In practice, this condition is realized when the nonlinear  pulses possess a larger group velocity than the linear surface modes for most  wavenumbers. Although our numerical simulations offer ample  evidence of the existence of nonlinear surface pulses, we have not yet been able to derive analytical solutions to prove their solitary nature. Given their characteristic width of about 5 units,   as well as their spatial shapes, the supported surface pulses  could be either compactons or micropterons \cite{Rosenau1993,Rosenau2018,Remoissenet1999}, but an analytical solution is needed to confirm this hypothesis - a challenge for future work.  

\begin{acknowledgments}
The authors gratefully acknowledge support via NSF award numbers 2041410 and 2041440. VT acknowledges support from project ExFLEM ANR-21-CE30-0003. JL acknowledges support from China Scholarship Council.
\end{acknowledgments}

\vspace{0.5cm}
{\textbf{Data availability.}} The data that support the findings of this study are available from the corresponding author upon reasonable request.

\bibliography{./APLmain.bib}

\end{document}